\documentstyle[12pt,epsfig]{article}

\def\lsim{\,\raise0.3ex\hbox{$<$\kern-0.75em\raise-1.1ex\hbox{$\sim$}}\,}
\def\gsim{\,\raise0.3ex\hbox{$>$\kern-0.75em\raise-1.1ex\hbox{$\sim$}}\,}

\newcommand{\comment}[1]{{}}

\begin{document}

\pagestyle{empty}
\begin{flushright}
HIP-2002-52/TH\\
CERN-TH/2002-322\\
hep-ph/0211239\\
November 2002\\
revised, March 2003
\end{flushright}
\vspace*{5mm}
\begin{center}

{\bf NONLINEAR CORRECTIONS TO THE DGLAP EQUATIONS \\ 
     IN VIEW OF THE HERA DATA}

\vspace*{0.5cm}
 K.J. Eskola$^{\rm a,b,}$\footnote{kari.eskola@phys.jyu.fi},
 H. Honkanen$^{\rm a,b,}$\footnote{heli.honkanen@phys.jyu.fi},
 V.J. Kolhinen$^{\rm a,b,}$\footnote{vesa.kolhinen@phys.jyu.fi},
 Jianwei Qiu$^{\rm c,}$\footnote{jwq@iastate.edu} and
 C.A. Salgado$^{\rm d,}$\footnote{carlos.salgado@cern.ch}

\vspace{0.3cm}
{\em $^{\rm a}$ Department of Physics, University of Jyv\"askyl\"a,\\
P.O.Box 35, FIN-40014 University of Jyv\"askyl\"a, Finland\\}
\vspace{0.1cm}
{\em $^{\rm b}$ Helsinki Institute of Physics,\\
P.O.Box 64, FIN-00014 University of Helsinki, Finland\\}
\vspace{0.3cm}
{\em $^{\rm c}$ Department of Physics and Astronomy, \\
Iowa State University, Ames, Iowa, 50011, U.S.A.\\}
\vspace{0.3cm}
{\em $^{\rm d}$ CERN, Theory Division, CH-1211 Geneva, Switzerland}
\vspace*{1cm}\\
{\bf Abstract} \\ \end{center}
\vspace*{3mm}
\noindent

The effects of the first nonlinear corrections to the DGLAP evolution
equations are studied by using the recent HERA data for the structure
function $F_2(x,Q^2)$ of the free proton and the parton distributions
from CTEQ5L and CTEQ6L as a baseline.  By requiring a good fit to the
H1 data, we determine initial parton distributions at
$Q_0^2=1.4$~GeV$^2$ for the nonlinear scale evolution.  We show that
the nonlinear corrections improve the agreement with the $F_2(x,Q^2)$
data in the region of $x\sim 3\cdot 10^{-5}$ and $Q^2\sim 1.5$~GeV$^2$
without paying the price of obtaining a worse agreement at larger
values of $x$ and $Q^2$.  For the gluon distribution the nonlinear
effects are found to play an increasingly important role at $x\lsim
10^{-3}$ and $Q^2\lsim10$~GeV$^2$, but rapidly vanish at larger values
of $x$ and $Q^2$. Consequently, contrary to CTEQ6L, the obtained gluon
distribution at $Q^2=1.4$~GeV$^2$ shows a power-like growth at small
$x$.  Relative to the CTEQ6L gluons, an enhancement up to a factor
$\sim6$ at $x=10^{-5}$, $Q_0^2=1.4$~GeV$^2$ reduces to a negligible
difference at $Q^2\gsim 10$~GeV$^2$.

\vfill\eject

\setcounter{page}{1}
\pagestyle{plain}

\section{Introduction}

Parton distribution functions (PDFs), $f_i(x,Q^2)$, are needed for the
computation of inclusive cross sections of hard, collinearly factorizable,
processes in
hadronic collisions.  At sufficiently large values of the interaction
scale $Q^2$ and the momentum fraction $x$, where power corrections are
negligible, the scale evolution of the PDFs is predicted quite
accurately by the Dokshitzer-Gribov-Lipatov-Altarelli-Parisi (DGLAP)
evolution equations \cite{DGLAP} derived from perturbative QCD.

In the global analyses of the PDFs of the free proton, such as in
Refs. \cite{mrst,cteq6}, the lowest order (LO), next-to-leading order
(NLO) and next-to-next-to-leading order (NNLO) parton distributions
are extracted within the DGLAP framework using constraints from the
measured cross sections of various hard processes and from the sum
rules.  In the proton case the procedure is well established: the
initial distributions given at some initial scale $Q_0^2$ are first
evolved to larger $Q^2$ by using the DGLAP equations, then a
comparison with the data is made over a wide range of $x$ and $Q^2$,
after which the initial distributions are iterated until a good global
fit to the data is obtained.  The initial distributions are thus the
non-perturbative input needed, the element that perturbative QCD
cannot predict. The data from deeply inelastic lepton-proton
scattering (DIS) play a key role in these analyses especially in the
region of the smallest values of $x$, where the DIS data from $ep$
collisions at DESY-HERA give the only constraints available.
Recently, the H1 collaboration at HERA \cite{H1} has measured the
structure function $F_2(x,Q^2)$ of the proton down to $x\sim 3\cdot
10^{-5}$ but still in the perturbatively accessible region $Q^2\ge
1.5$ GeV$^2$. These data have been included in the recent global
analyses by the MRST \cite{mrst} and CTEQ \cite{cteq6} collaborations.

In spite of the impressive success of the DGLAP approach, certain
problems appear in the attempts to make the global fits to the H1 data
\cite{H1} as good as possible {\em simultaneously} in the region of
$Q^2>4$ GeV$^2$ (``large'' $Q^2$) and in the region of $1.5\,{\rm
GeV}^2<Q^2<4$ GeV$^2$ (``small'' $Q^2$). In the recent NLO analysis
MRST2001 \cite{mrst} both regions are included and
 a good overall fit is found but with the expense
of allowing for a negative gluon distribution.  Although a negative
contribution in the NLO gluon distribution is acceptable as long as
the NLO cross sections remain positive, the interpretation of the PDFs
as probability or number density distributions becomes obscured
\footnote{See also \cite{BRODSKY} for a recent discussion of the PDFs
not being probabilities.}. On the other hand, the CTEQ collaboration
emphasizes the large-$Q^2$ region in their global fits: e.g. in the
sets CTEQ5 \cite{cteq5} and CTEQ6 \cite{cteq6} only the region
$Q^2>$~4 GeV$^2$ of the DIS data is included in the fit and a very good
agreement with the data is found.  The agreement of the extrapolation
to the small-$Q^2$ region, however, becomes then worse. 
One is facing the problem of negative gluon distributions also in the
NLO set CTEQ6M \cite{cteq6}: $xg(x,Q^2)$ is set to zero at the
smallest values of $x$ at scales $Q^2\lsim1.69$~GeV$^2$.  In LO, the
negative gluon distributions, however, are not allowed. The quality of
the LO fits to the H1 data (see Table~1 in Sec.~3) reflects the problem
of a simultaneous fit to the small- and large-$Q^2$ regions: MRST2001
(CTEQ6) fits the small-$Q^2$ (large-$Q^2$) region better.

The problems discussed above are very interesting as they can be a
sign of a new QCD phenomenon: towards smaller values of $x$ and (or)
$Q^2$ (but still $Q^2\gg\Lambda_{\rm QCD}^2$), gluon recombination
effects are expected to play an increasingly important role.  These
effects induce nonlinear power corrections to the DGLAP equations.
First of the nonlinear terms have been calculated by Gribov, Levin and
Ryskin in \cite{GLR} and by Mueller and Qiu in \cite{MQ}. We shall
refer to these corrections as the GLRMQ terms. 

Previous studies of the GLRMQ terms in the context of extracting the
PDFs of the free proton can be found e.g. in \cite{kmrs}.  Also other
nonlinear evolution equations relevant at high gluon densities have
been derived in the recent years \cite{eveqs}, and the structure
functions from DIS have been analysed in the context of saturation
models \cite{phenan}. In the present work, however, we shall adopt the
framework of collinear factorization with universal PDFs, and search
for the nonlinear GLRMQ corrections on top of the full DGLAP
equations. This allows for a direct comparison of our results with
those of the global DGLAP fits \cite{mrst,cteq6}. In this manner we
can also show more explicitly the need for nonlinear terms in the
evolution equations.

In the DGLAP evolution, the $Q^2$ dependence of the sea quarks at
small values of $x$ is dictated by the gluon distribution: the larger
$xg(x,Q^2)$, the faster the $Q^2$ evolution of $F_2(x,Q^2)$, since in
LO $\partial F_2(x,Q^2)/\partial \log Q^2 \approx (10\alpha_s/27\pi)
xg(2x,Q^2)$ \cite{PRYTZ}. The GLRMQ terms slow down the $Q^2$ evolution of
gluons and sea quarks from the standard DGLAP behaviour
(assuming the same starting distributions). Consequently, the $\log
Q^2$ slopes of $F_2(x,Q^2)$ of the H1 data can be reproduced with
a {\em larger} gluon distribution than that in the
conventional DGLAP case. In this paper, we investigate this
interdependence of the initial conditions at $Q_0^2=1.4$~GeV$^2$ and
the effects of the GLRMQ terms by using the H1 data as a baseline.
We demonstrate that inclusion of the GLRMQ terms on top of the LO
DGLAP evolution improves the agreement with the data in the small-$x$
and small-$Q^2$ region while simultaneously maintains the
good fit of the LO sets of CTEQ5 and CTEQ6 to the data at
larger values of $x$ and $Q^2$. 
We show that the obtained small-$x$ gluon distribution can still have
a power-like growth at the scale $Q^2=1.4$~GeV$^2$, leading to an
enhancement of a factor $\sim6$ relative to the CTEQ6L gluons at
$x=10^{-5}$. We also show explicitly how the large deviations from
CTEQ6L reduce to very small differences at $Q^2\gsim10$~GeV$^2$.  The
size of the nonlinear terms, uncertainties and applicability region of
the DGLAP+GLRMQ approach are discussed as well.

\section{Nonlinear evolution equations}

The GLRMQ corrections \cite{GLR,MQ} arise from fusion of two gluon
ladders, and they modify the evolution equations of gluons as

\begin{equation}
\frac{\partial xg(x,Q^2) }{\partial \log Q^2} 
=   \frac{\partial xg(x,Q^2) }{\partial \log Q^2}\bigg|_{\rm DGLAP}
  - \quad \frac{9\pi}{2} \frac{\alpha_s^2}{Q^2} 
    \int_x^1 \frac{dy}{y} y^2 G^{(2)}(y,Q^2),
\label{MQgluon}
\end{equation} 
where the first term is the standard DGLAP result \cite{DGLAP}, linear in the
PDFs.  The 2-gluon density in the second term we model as
\begin{equation}
x^2G^{(2)}(x,Q^2)= \frac{1}{\pi R^2}[xg(x,Q^2)]^2, 
\label{2gluondensity}
\end{equation}
with $R=1$~fm as
the radius of the proton. The equation (\ref{MQgluon}) thus becomes
nonlinear in $xg$. In the scale evolution of the sea quark
distributions the leading nonlinear correction from the gluon fusion 
appears as \cite{MQ}

\begin{equation}
\frac{\partial xq(x,Q^2)}{\partial \log Q^2}   \approx  
\frac{\partial xq(x,Q^2)}{\partial \log Q^2}\bigg|_{\rm DGLAP}\nonumber\\
 -  \quad \frac{3\pi}{20}\frac{\alpha_s^2}{Q^2} 
     x^2 G^{(2)}(x,Q^2), 
\label{MQsea}
\end{equation} 
where the first term is again from the DGLAP equations and linear in
the PDFs.  
Note that here we work under an approximation of neglecting
contribution from the ``high twist'' gluon distribution $G_{\rm HT}(x,Q^2)$ 
\cite{MQ}.  More discussion on this will be given later \cite{JQprep}.
The evolution of the valence quark distributions remains unmodified.

\section{Analysis and Results}

\subsection{Effects of the GLRMQ corrections}

The emphasis of the CTEQ analysis is in the large-$Q^2$ region where
the nonlinearities should remain small. Since we wish to search for
the nonlinear effects in the small-$Q^2$ (small-$x$) region but
also recover the DGLAP evolution at large $Q^2$, the CTEQ distributions 
are an ideal baseline for our analysis.


\begin{figure}[tb]
\vspace{-0.5cm}
\centering{\epsfxsize=15cm\epsfbox{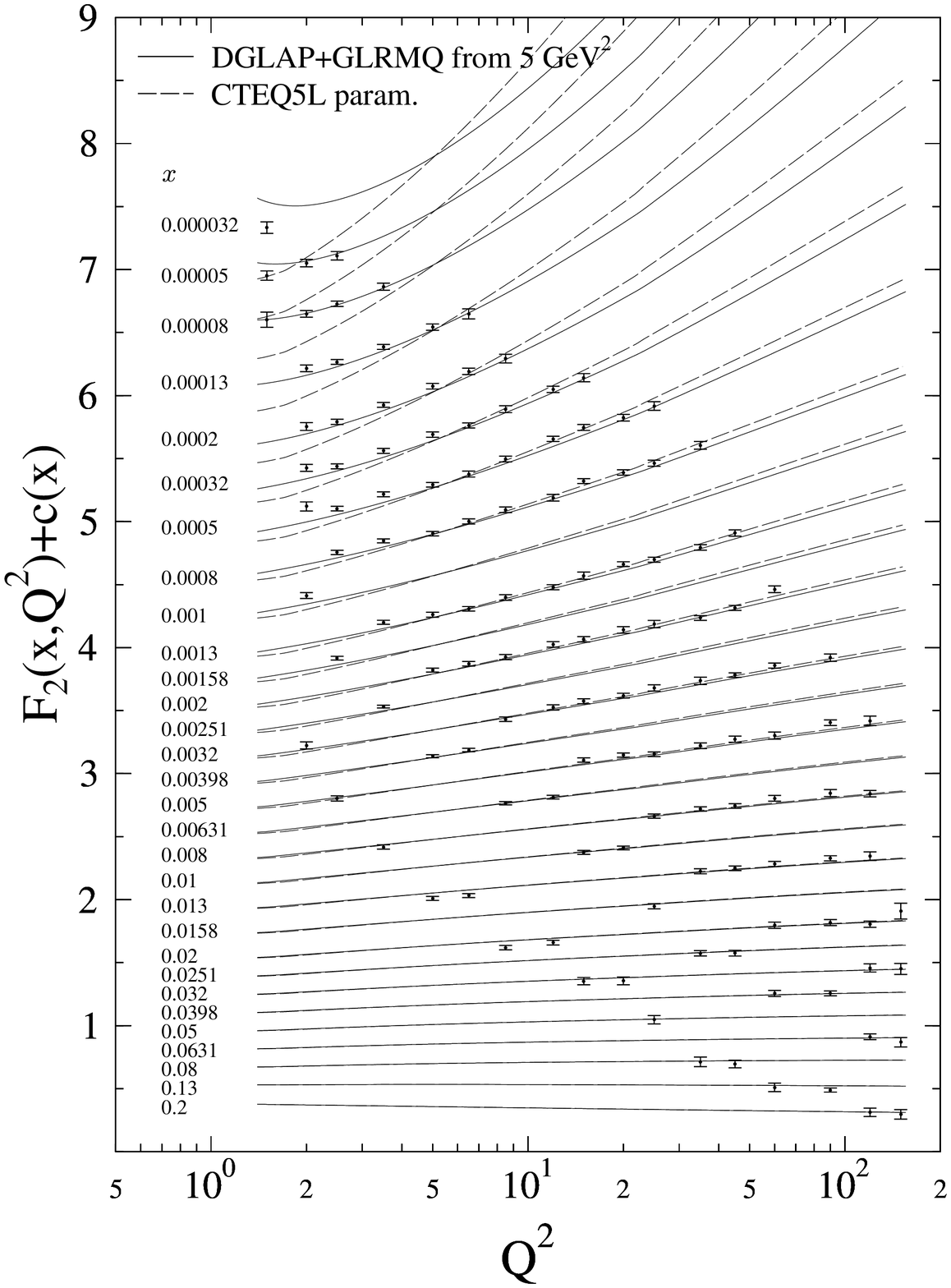}}
\vspace{0cm}
\caption[a]{{\small The scale evolution of the structure function
$F_2(x,Q^2)$ of the free proton for fixed values of $x$ (with
constants added to separate the curves).  The dashed curves show the
LO DGLAP result from CTEQ5L \cite{cteq5}, and the solid curve the
result after the DGLAP+GLRMQ evolution when initial conditions taken
from CTEQ5L at $Q_0^2=5$~GeV$^2$. The data is from H1 \cite{H1} and
the error bars are statistical.} }
\vspace{0cm}
\label{F2_vs_cteq5}
\end{figure}

\begin{figure}[tb]
\vspace{-0.5cm}
\centering{\epsfxsize=13cm\epsfbox{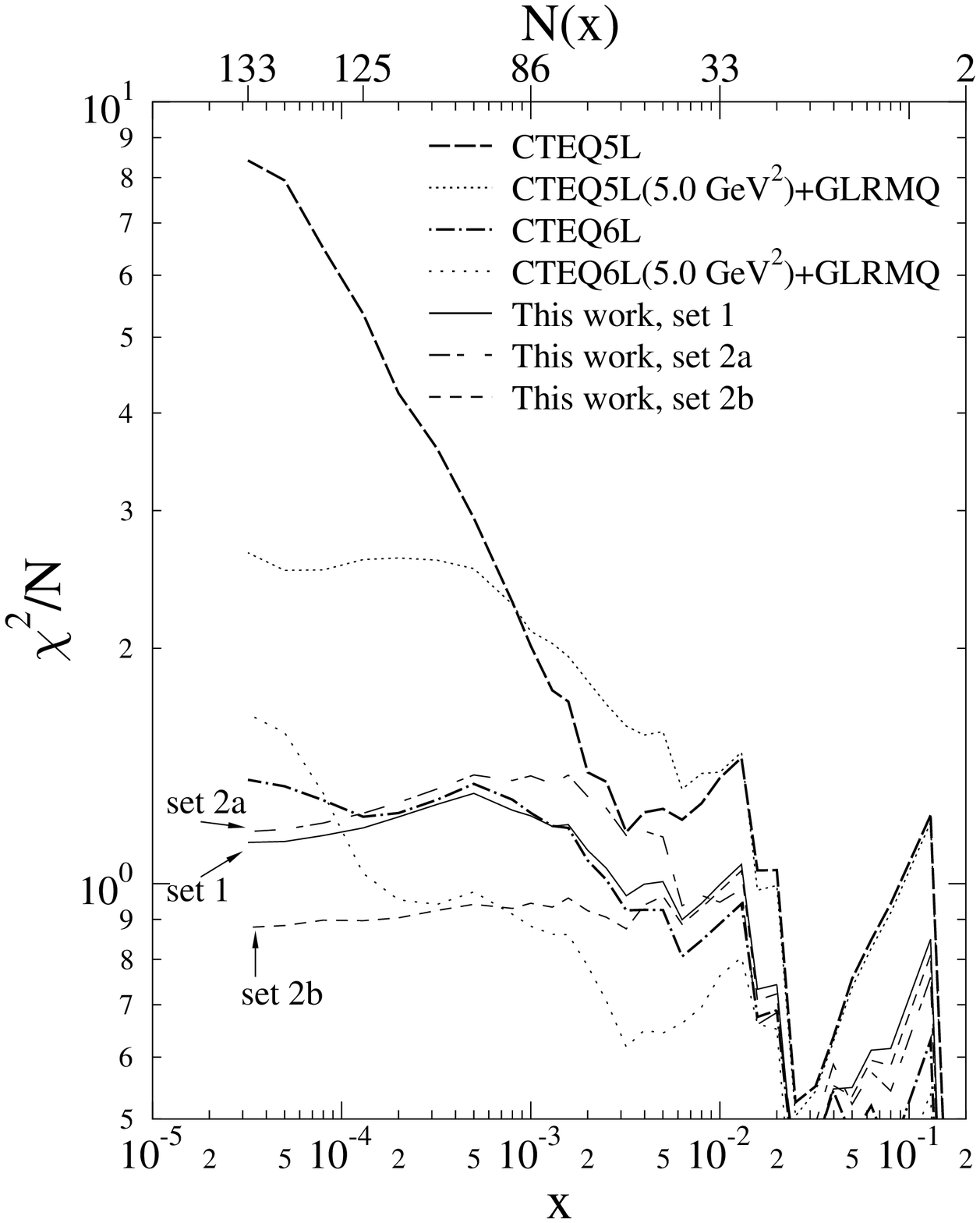}}
\vspace{0cm}
\caption[a]{{\small The goodness parameter $\chi^2$ of the fits of the
computed $F_2(x,Q^2)$ to the H1 data, divided by the number of data
points, as a function of the $x$ of the data. The cumulative number of
the data points is increasing to the left as indicated at the top of
the plot: $N(x=0.2)=2$ and $N(x=3.2\cdot10^{-5})=133$ (see also Table
1).  The curves are the LO DGLAP results from CTEQ5L (long dashed
thick line) and CTEQ6L (dotted-dashed thick line), the DGLAP+GLRMQ
result with the initial conditions at $Q^2=5$~GeV$^2$ taken from
CTEQ5L (densely dotted) and from CTEQ6L (sparsely dotted), and, our
set 1 (solid), set 2a (double dashed) and set 2b (short dashed) .} }
\vspace{0cm}
\label{khi2}
\end{figure}

In Fig. \ref{F2_vs_cteq5}, together with the H1 data \cite{H1}, we
plot the $Q^2$ dependence of the LO structure function
$F_2(x,Q^2)=\sum_q e_q^2[xq(x,Q^2)+x\bar q(x,Q^2)]$ computed from
CTEQ5L \cite{cteq5,PDFLIB} for fixed values of $x$ (dashed lines).
The agreement with the H1 data is clearly getting worse towards
smaller values of $x$ and $Q^2$. This trend is clearly visible also in
Fig. \ref{khi2} (dashed line), where we quantify the quality of the
fit in terms of
\begin{equation}
\chi^2(x_k)=\sum_{x_j=0.2}^{x_k} 
\sum_{i=1}^{n(x_j)}
\frac{ [F_2^{\rm th}(x_j,Q_i^{(x_j)})
      -F_2^{\rm exp}(x_j,Q_i^{(x_j)})]^2}
{[\Delta_{F_2}^{\rm exp}(x_j,Q_i^{(x_j)})]^2}
\end{equation}
 divided by the cumulative number of data points, $N(x_k)=
\sum_{x_j=0.2}^{x_k}n(x_j)$, as a function of the $x$ of the data.
Here $n(x_j)$ refers to the number of the data points with the same
value of $x=x_j$.  The $\chi^2/N$ computed at different regions of
$Q^2$ is in turn summarized in Table 1.  Relative to the data
\cite{H1} (which was not available at the time of CTEQ5) the computed
$\log Q^2$ slopes of $F_2$ are too large at small values of $x$ and
$Q^2$. This is caused by a too large gluon component in the DGLAP
evolution at small values of $x$.


\begin{table}[tb]
\begin{center}
\begin{tabular}{|l|c|c|c|}
\hline 
 	 & $Q^2<4.0$ GeV$^2$ 	& $Q^2>4.0$ GeV$^2$  	& all $Q^2$ 	\\  
	 & $N=29$		& 	$N=104$		& 	$N=133$	\\
\hline  
~~~~~~~~~CTEQ5L	 
	 &	31.8		&	1.18		& 	7.86	\\
~~~~~~~~~CTEQ6L
	 &	2.72		&	0.93		&	1.32	\\
~~~~~~~~MRST2001
	 &	0.59		&	2.06		&	1.74	\\
This work:&			&			&		\\ 
Set 1:  $Q_c< \sqrt{1.4}$~GeV
	 &	1.75		&	0.96		&	1.13	\\
Set 2a: $Q_c=1.3$~GeV
	 &	1.58		&	1.05		&	1.17	\\
Set 2b: $Q_c=\sqrt{1.4}$~GeV
	 &	0.95		&	0.86		&	0.88	\\

\hline
\end{tabular}
\end{center}
\caption{\small The $\chi^2/N$ of the fit of the computed LO $F_2$ to
the H1 data \cite{H1} in the small-$Q^2$ and large-$Q^2$ regions. 
Also the number of data points in each case is mentioned.}
\end{table}

For showing how the GLRMQ terms slow down the scale evolution, we take
the PDFs from the CTEQ5L parametrization at $Q^2=5$~GeV$^2$, and
evolve these both downwards and upwards by using the nonlinear
evolution equations. The results are plotted in Fig. \ref{F2_vs_cteq5}
(solid lines). As seen in the figure, at larger values of $x$ and
$Q^2$ the nonlinear effects remain small and do not make the agreement
with the data worse. In addition, as shown both by
Fig. \ref{F2_vs_cteq5} and Fig. \ref{khi2}, at $5\cdot10^{-5}\lsim
x\lsim 10^{-3}$ the agreement with the data becomes quite good
(i.e. $\chi^2/N$ remains constant), clearly improving the situation from
the CTEQ5L case.

Another interesting observation from Fig. \ref{F2_vs_cteq5} is that,
with the trial initial conditions above, the nonlinear corrections at
$x\lsim5\cdot10^{-5}$ and $Q^2\sim 1$~GeV$^2$ obviously become too
large and they start to dominate the computed evolution, causing
negative $\log Q^2$ slopes for $F_2(x,Q^2)$. Clearly, these are not 
supported by the data.
In the computation we would then have entered a gluon saturation
region where also other correction terms in the evolution equations
should be included.  This region is thus beyond the scope of the
present DGLAP+GLRMQ approach. The too strong nonlinearities 
can again be traced back to a too large gluon component at small $x$ in
the initial condition chosen.  A more realistic gluon
distribution which evolves according to the DGLAP+GLRMQ equations
should obviously be smaller than CTEQ5L
at small values of $x$ at $Q^2=5$~GeV$^2$.

\begin{figure}[tb]
\vspace{-0.5cm}
\centering{\epsfxsize=15cm\epsfbox{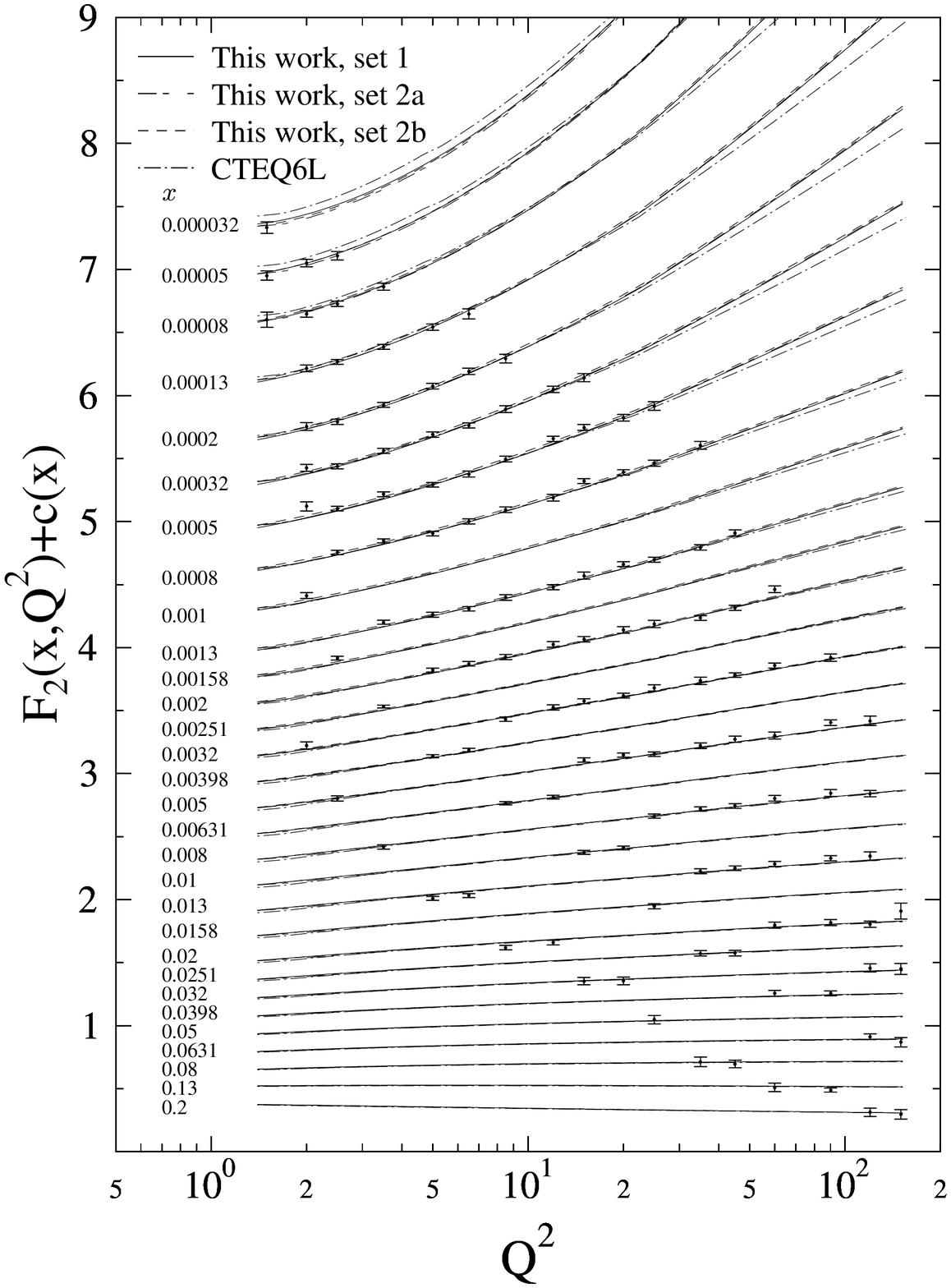}}
\vspace{0cm}
\caption[a]{{\small As Fig.~\ref{F2_vs_cteq5} but for LO DGLAP result
from CTEQ6L (dotted-dashed) and for the DGLAP+GLRMQ results with our
set 1 (solid), set 2a (double dashed) and set 2b (short dashed).} }
\vspace{0cm}
\label{F2_vs_cteq6}
\end{figure}

The H1 data \cite{H1} is partly taken into account in the latest set
CTEQ6L \cite{cteq6}, and the situation is clearly improved from
CTEQ5L. This is shown in Figs.  \ref{F2_vs_cteq6} and \ref{khi2} by
the dotted-dashed curves.  In Fig. \ref{xf_panels} we have plotted the
PDFs at $Q^2=1.4$~GeV$^2$ for several cases. Now, as shown by
Fig. \ref{xf_panels}, the gluon distributions (upper left
panel) of CTEQ6L (dotted-dashed) at small $x$ are smaller than those
of CTEQ5L (dashed), causing more modest $\log Q^2$ slopes for $F_2$ at
small $x$. Fig.  \ref{khi2} indicates that the agreement of CTEQ6L
with the H1 data is excellent at $x\gsim 10^{-3}$ and stays very good
also at smaller $x$.  Table 1 again expresses $\chi^2/N$ at the
large-$Q^2$ region (included in the CTEQ6 global fit) and in the
small-$Q^2$ region (not included in the CTEQ6 fit). Although a good
agreement with the data is found, a closer look at Fig.
\ref{F2_vs_cteq6} shows that again the agreement of the pure DGLAP
result is getting worse towards the smallest values of $x$.

We notice that, contrary to CTEQ5L, the CTEQ6L result in
Fig. \ref{F2_vs_cteq6} lies above the H1 data at the smallest values
of $x$. If we now simply took the initial conditions from CTEQ6L at,
say, $Q^2=5$~GeV$^2$, and evolved the distributions downwards by using
the nonlinear equations (which would slow down the evolution), we
would make the agreement with the data in the small-$x$ region {\em
worse} than the CTEQ6L result. The dotted curve in Fig. \ref{khi2}
shows the $\chi^2/N$ for such a run, notice the growing trend towards
the smallest $x$. Encouraged, however, by the observations with the
CTEQ5L+GLRMQ above, we wish to see whether we could find initial
conditions at some $Q_0^2$ that would lead to at least the same or
possibly even better agreement with the H1 data as the DGLAP result
from CTEQ6L.

\subsection{New initial distributions}

\subsubsection{Set 1}


\begin{figure}[tb]
\vspace{-0.5cm}
\centering{\epsfxsize=15cm\epsfbox{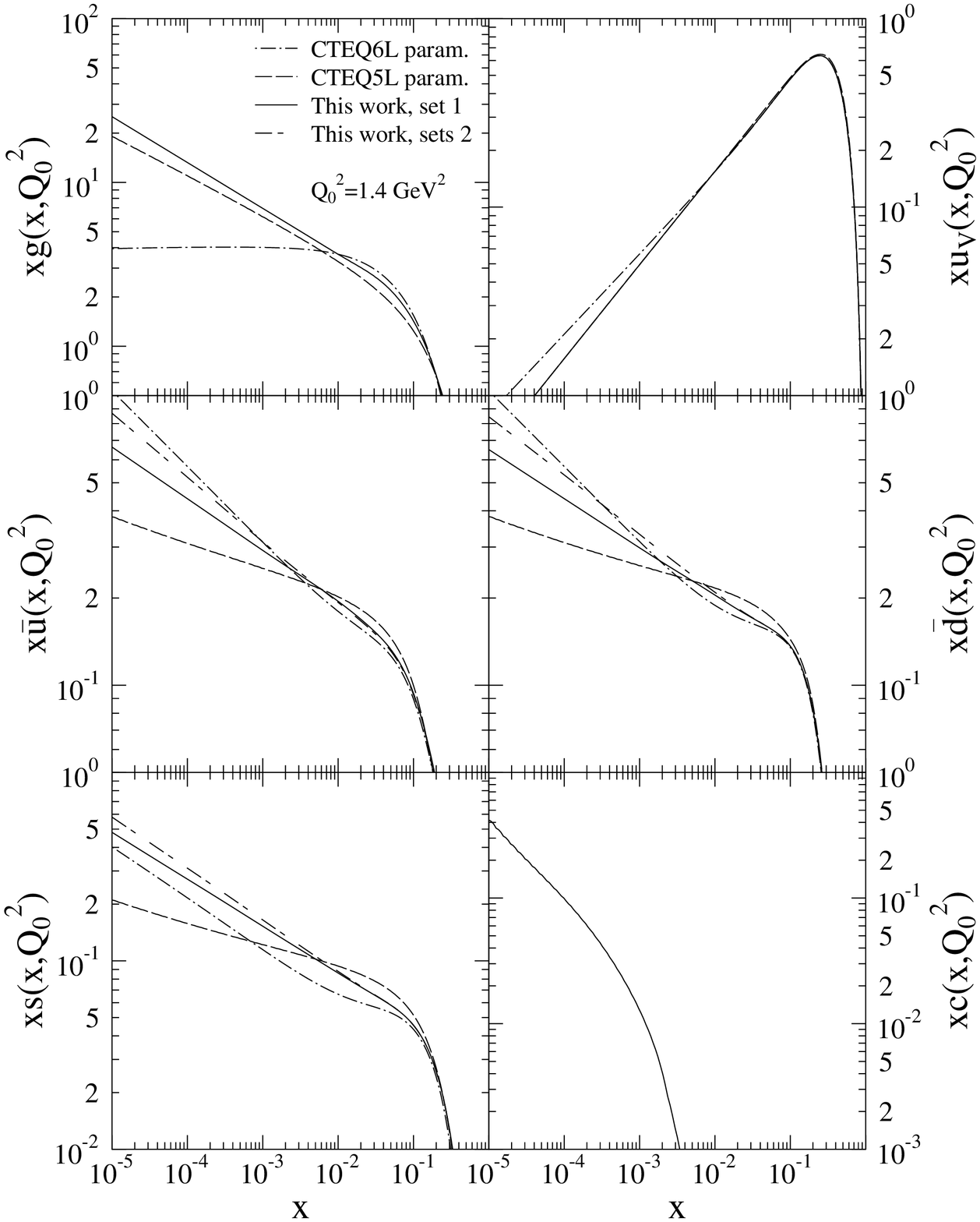}}
\vspace{0cm}
\caption[a]{{\small The parton distribution functions at
$Q^2=1.4$~GeV$^2$ as obtained in the DGLAP analyses CTEQ5L
\cite{cteq5} (dashed), CTEQ6L \cite{cteq6} (dotted-dashed) and in the
present work based on the DGLAP+GLRMQ evolution. In our set 1 (solid),
there is a finite charm contribution, while in the sets 2 (double
dashed) charm is zero at this scale.  Notice the enhancement from the
CTEQ6L glue. The gluon distributions of set 1 and sets 2 are
identical.}}
\vspace{0cm}
\label{xf_panels}
\end{figure}

We construct the initial distributions at an initial scale
$Q_0^2=1.4$~GeV$^2$ by using the CTEQ5L and CTEQ6L sets as our
guide. Following CTEQ5L we use $\Lambda_{\rm QCD}^{(4)}=192$~MeV
together with the one-loop expression of the strong coupling constant.
The change of the value of $\Lambda_{\rm QCD}$ at the heavy-quark mass
thresholds is taken into account. Throughout the study we use
$Q_b=4.75$~GeV for the $b$-threshold, and the $c$-threshold will be
discussed below. As seen in Fig. \ref{khi2} (sparsely dotted curve), a
good agreement with the H1 data is found at $x\gsim0.01$ by taking the
CTEQ6L PDFs at 5.0 GeV$^2$ and evolving them down to 1.4~GeV$^2$
according to the nonlinear equations. The CTEQ5L distributions evolved
down to 1.4~GeV$^2$ from 10.0 GeV$^2$ and 3.0 GeV$^2$ with the GLRMQ
corrections included, give a reasonably good agreement with the H1
data at $10^{-4}\lsim x \lsim 0.01$ and at $10^{-5}\lsim x\lsim
10^{-4}$, correspondingly (not shown).  A working initial condition
can then be found by interpolating between these three
results. Finally, we make a power-law fit to the interpolated gluon
and light sea-quark distributions at small values of $x$. The
$x$-slope of the small-$x$ gluons is tuned to reproduce the measured
$\log Q^2$ slopes of $F_2$. This leads to an initial gluon
distribution $xg(x,Q_0^2)=3.64\cdot(0.01/x)^{0.28}$ at small $x$.

The initial conditions constructed in this way at $Q_0^2=1.4$~GeV$^2$
are shown in Fig. \ref{xf_panels}, labelled as our ``set 1''.  The
obtained distributions are compared with the CTEQ5L and CTEQ6L PDFs at
the same scale. Notice especially that our gluon distribution at
$x=10^{-5}$ at this scale is a factor $6.4$ larger than the one in
CTEQ6L.  In turn, the light sea-quark distributions at small $x$
lie in between CTEQ6L and CTEQ5L.  Due to the procedure we have chosen,
the valence-quark distributions (which evolve according to DGLAP) are
practically the same as in CTEQ5L at $x\lsim0.01$ and CTEQ6L at
$x\gsim0.01$.

The DGLAP+GLRMQ evolution to higher scales then gives the result shown
in Fig. \ref{F2_vs_cteq6} by the solid lines.  The corresponding
values of $\chi^2/N$ are again shown as a function of $x$ in
Fig. \ref{khi2} and its division into small-$Q^2$ and large-$Q^2$
regions in Table 1. We observe that while the agreement with the H1
data at large values of $x$ and $Q^2$ is maintained practically as
good as in CTEQ6L, the fit in the smallest-$x$, small-$Q^2$ region is
indeed improving. This, together with the parton distributions at
$Q_0^2=1.4$~GeV$^2$ for the nonlinear evolution, is the main result of
this paper.  At this point we should also emphasize that we have not
attempted to make a global statistical analysis such as the one by
CTEQ in order to further minimize the $\chi^2$. After such a
procedure, further improvement on the $\chi^2$ could well be
anticipated.

\begin{figure}[tb]
\vspace{-0.5cm}
\centering{\epsfxsize=15cm\epsfbox{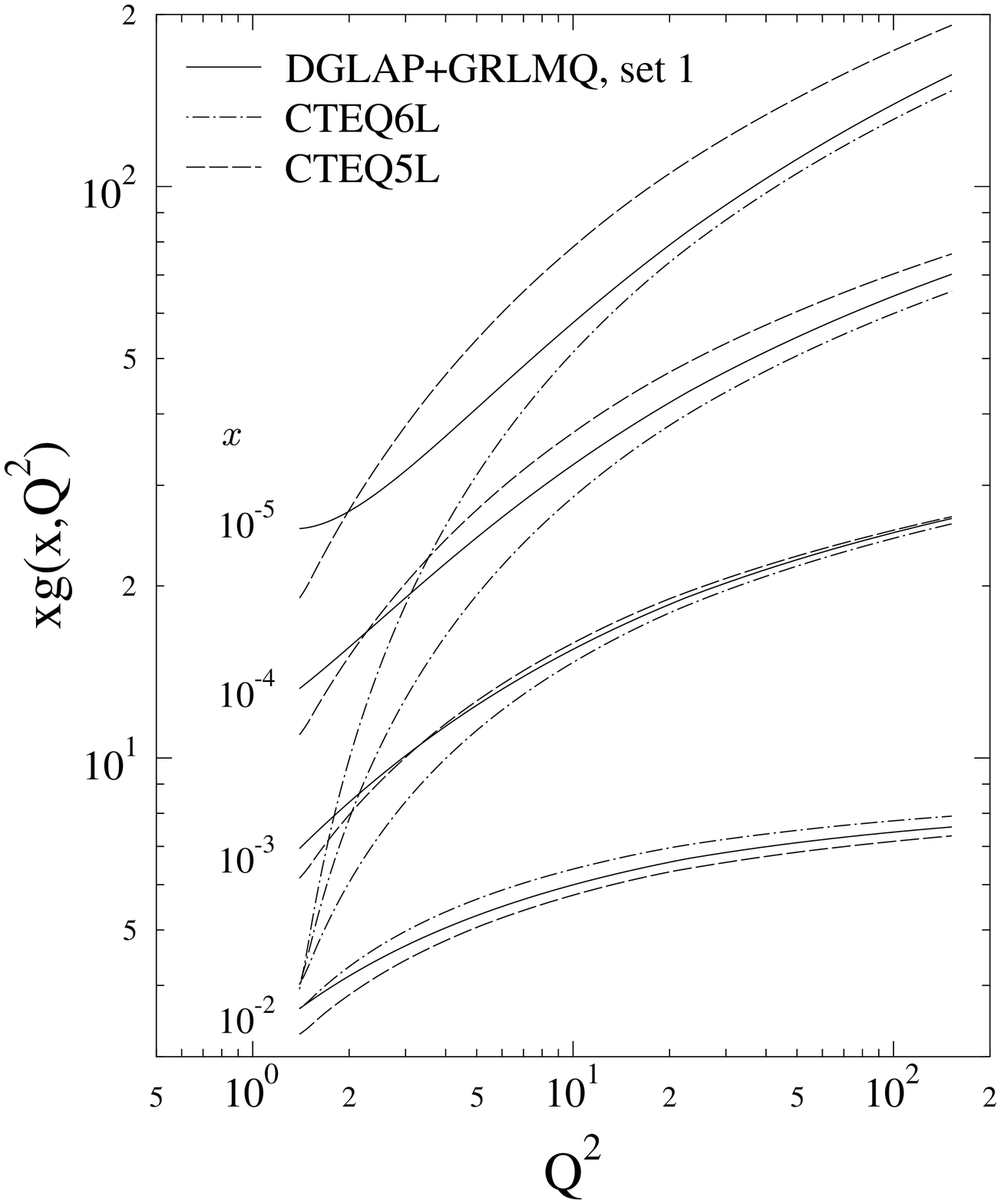}}
\vspace{0cm}
\caption[a]{{\small The $Q^2$ dependence of the gluon distribution 
function at fixed values of $x$, from  CTEQ5L \cite{cteq5} (dashed),
CTEQ6L \cite{cteq6} (dotted-dashed) and set 1 of the present work
(solid). Notice the logarithmic scales and the absolute normalization 
of the curves.}}
\vspace{0cm}
\label{gluons}
\end{figure}

Fig.~\ref{gluons} shows the $Q^2$ dependence of the gluon
distributions at fixed values of $x$ as obtained in the DGLAP analyses
CTEQ5L (dashed lines) and CTEQ6L (dotted-dashed) as well as from the
GLRMQ+DGLAP evolution of the gluons of our set 1 (solid).  Notice on
one hand the large difference in the CTEQ5L and CTEQ6L sets, and on
the other hand the slower evolution in the nonlinear case at small
values of $x$ and $Q^2$. We also draw attention to the fact that in
spite of the large (factor 6.4) difference at $Q^2_0=1.4$~GeV$^2$ at
$x=10^{-5}$, our gluon distributions and the CTEQ6L gluons are in fact
quite similar at $Q^2\gsim 5$~GeV$^2$. Coming back to the discussion
related to Fig~\ref{F2_vs_cteq5}, we also observe from
Fig.~\ref{gluons} that at $Q^2=5$~GeV$^2$ the gluons from our set 1
are below the CTEQ5L gluons as anticipated.

\subsubsection{Sets 2a and 2b}

A detail to study next is the $c$-quark contribution. The initial
distributions in our set 1 were obtained on the basis of distributions 
which were evolved downwards from 3, 5 and 10 GeV$^2$
according to the nonlinear evolution equations. The $c$-quarks we
treat as massless quarks which experience the GLRMQ corrections as
well. Therefore, the downwards evolution of the $c$-quarks is also
slower than in the DGLAP case, and their distributions vanish only
somewhat below the threshold scale $Q_c=1.3$~GeV of the sets CTEQ6L 
and CTEQ5L. This explains the fairly large $c$-distribution at $Q_0^2$ 
in our set 1.

We construct two slightly different sets of PDFs with the same initial
conditions at $Q_0^2=1.4$~GeV$^2$, where $xc(x,Q_0^2)=0$. For the
first case, called ``set 2a'' we follow CTEQ6L and take
$Q_c=1.3$~GeV as the $c$-threshold scale.  For the other case,
called ``set 2b'' we choose, $Q_c=Q_0=\sqrt{1.4}$~GeV.  In these
initial conditions, the gluons are not modified from set 1. However,
in order to compensate for the loss of the initial $c$-quarks and to
recover the agreement with the H1 data, we enhance the other
sea-quarks in the small-$x$ region simply by tuning their power in the
power-law fit. The initial distributions for our sets 2 are shown in
Fig.~\ref{xf_panels} (double dashed).  After the nonlinear evolution
to higher scales, the agreement of the set 2a with the H1 data is
practically as good as with our set 1 and CTEQ6L, as is shown in
Figs.~\ref{F2_vs_cteq6} and \ref{khi2} (double dashed lines). With the
set 2b, the agreement becomes even better, as can be seen from 
Figs.~\ref{khi2} and \ref{F2_vs_cteq6}  (short dashed lines).
The same conclusion is suggested also by the
$\chi^2/N$ computed in the different regions of $Q^2$ in Table 1. The
GLRMQ corrections thus improve the fits to the data in the region of small
$x$ and $Q^2$ without loosing the good fits at larger $x$ and
$Q^2$. Regarding the sensitivity of the results to the $c$-threshold,
we note that the differences between our sets 2a and 2b are quite
small and could most probably be obtained also by keeping the
$c$-threshold constant while tuning the gluon distributions at
$0.001\lsim x\lsim 0.01$. This fine-tuning is, however, beyond the
goal of this paper.

\section{Discussion and Conclusions}

It is important to keep in mind the uncertainties and limitations of
the present approach. A constraint not addressed above but used in the
global analysis of the PDFs, is momentum conservation. Due to the
nonlinear terms in Eqs. (\ref{MQgluon}) and (\ref{MQsea}), some
momentum from small values of $x$ is lost in the evolution towards higher
scales.  Due to this, our sets 1 and 2 overestimate the total momentum
at $Q_0^2=1.4$~GeV$^2$ by less than 2 \%. By $Q^2=100$~GeV$^2$, some 2.6 \%
of the initial total momentum is lost.  As the emphasis of our study
is in the small-$x$ behaviour of the PDFs, and as the violation
remains small, we have not made an attempt to correct the obtained
distributions for the momentum sum rule.

In modelling the 2-gluon density in Eq. (\ref{MQgluon}), we have set
the effective radius parameter of the free proton to $R=1$~fm.
Depending on the transverse matter density profile assumed for the
free proton, some $\sim$ 20 \% uncertainty in $R$ can be expected. The
nonlinearities decrease with increasing $R$, so due to the interplay
between the initial conditions and the scale evolution demonstrated
above, a larger $R$ would lead to a smaller enhancement of the
small-$x$ gluons.  In order to properly estimate the uncertainty in
the initial gluon distribution caused by the uncertainty in $R$, the
fit analysis performed above should be redone with modified values of
$R$. While this is beyond the scope of the present paper, it is left
as a future task.

The form $\sim (xg)^2$ for the 2-gluon density also neglects
possible longitudinal correlations which should suppress the
2-gluon density at sufficiently large values of $x$. Also the possible
difference of the fractional momenta of the fusing gluons is
neglected.  It can be argued, based on a simple picture of a Lorentz
contracted proton and wavelengths of partons given by their inverse
momentum, that gluons with $x<1/(2m_pR)\sim 0.1$ overlap with any
other gluons and thus cause finite nonlinear corrections.  As the 1-gluon
densities are already decreasing quite rapidly at $x\gsim 0.1$, and as
their $\log Q^2$ slopes become anyway small there, we have not attempted
to build in any longitudinal correlations of the fusing gluons. In a
larger system, in a big nucleus, this would be necessary, and
sensitivity to such details should be studied. 

Another obvious improvement for the present analysis is to consider the
GLRMQ terms added on top of the NLO DGLAP evolution. In that case, the
zero (CTEQ6M) or negative (MRST2001) gluon distributions at small $x$
may become larger than zero.

In the present study we have neglected a possible but small
contribution from the higher-dimensional gluon distribution $G_{\rm
HT}$ introduced in \cite{MQ}. The distribution $G_{HT}$ can be thought
as a $k_T^2$ moment of $k_T$-dependent gluon distribution, and its
upper limit estimated as $G_{HT}(x,Q^2) \approx \langle k_T^2\rangle
g(x,Q^2) < Q^2 g(x,Q^2)$.  Therefore, this term should be less than
the normal DGLAP term, and remain negligible. Further studies on this
interesting question will follow \cite{JQprep}.

Regarding the size of the nonlinear terms, it is quite interesting to
notice that in our results at $Q_0^2=1.4$~GeV$^2$, the GLRMQ terms in
Eq. (\ref{MQgluon}) make about 48 \% of the full DGLAP+GLRMQ $\log
Q^2$ slope of the gluon distribution at $x=10^{-5}$, and still some
16\% at $x=0.01$.  The extent of nonlinearity decreases, as expected,
with increasing $Q^2$: at $Q^2=10$~GeV$^2$ the GLRMQ contribution to
the total $\partial xg/\partial \log Q^2$ is 26 \% at $x=10^{-5}$ and
below 4\% at $x=0.01$.  At the lowest values of $Q^2$ and $x$ probed
by the H1 data \cite{H1}, we clearly are at the borderline of the
applicability of the approach, i.e. close to the gluon saturation
region, where the next terms in the nonlinear evolution equations are
becoming important \cite{AYALA,Jalilian,Weigert,Kovchegov}.  The
further nonlinear correction terms obviously enter with an alternating
sign. In the region where the GLRMQ term in Eqs. (\ref{MQgluon}) and
(\ref{MQsea}) becomes as important as the DGLAP term, inclusion of the
further corrections should thus decrease the net
correction. Therefore, the results of the current paper in the region
of smallest $x$ and $Q^2$ studied can be regarded as an upper limit of
the small-$x$ gluon distributions.

In conclusion, we have studied the effects of adding the nonlinear
GLRMQ corrections to the LO DGLAP evolution equations, and especially
the interplay between the initial conditions and the
nonlinearities. We use the PDF sets CTEQ5L and CTEQ6L as a baseline,
and the recent DIS data from H1 \cite{H1} as a constraint.  We have
shown that the agreement between the measured and computed structure
function $F_2(x,Q^2)$ can be improved at small values of $x$ and $Q^2$
while still maintaining the good fit to the data obtained in the
global analyses at larger values of $x$ and $Q^2$.  The nonlinearities
slow down the scale evolution, so in order to recover the measured
$\log Q^2$ slopes of $F_2(x,Q^2)$ of the data, a larger small-$x$
gluon distribution than that in CTEQ6L is needed. For the gluon
distribution the nonlinear effects are found to play an increasingly
important role at $x\lsim 10^{-3}$ and $Q^2\lsim10$~GeV$^2$. The
nonlinearities, however, vanish rapidly at larger values of $x$ and
$Q^2$. Consequently, contrary to CTEQ6L, the obtained gluon
distribution at $Q^2=1.4$~GeV$^2$ shows a power-like growth at small
values of $x$.  Relative to the CTEQ6L gluons, an enhancement up to a
factor $\sim6$ at $x=10^{-5}$, $Q^2=1.4$~GeV$^2$ reduces to a
negligible difference at $Q^2\gsim 10$~GeV$^2$.

\vspace{1cm}
\noindent{\large \bf Acknowledgements.}  We thank N. Armesto,
P.V. Ruuskanen, I. Vitev and other participants of the CERN Hard
Probes workshop for discussions. We are grateful for the Academy of
Finland, Project 50338, for financial support.  J.W.Q. is supported in
part by the United States Department of Energy under Grant
No. DE-FG02-87ER40371.  C.A.S. is supported by a Marie Curie
Fellowship of the European Community programme TMR (Training and
Mobility of Researchers), under the contract number
HPMF-CT-2000-01025.

\end{document}